# Twofolds in C and C++

Evgeny Latkin
2014 Dec 13

https://sites.google.com/site/yevgenylatkin/
Email: yevgeny.latkin@gmail.com

**Abstract**: Here I propose C and C++ interfaces and experimental implementation for twofolds arithmetic I introduce in [1] for tracking floating-point inaccuracy. Testing shows, plain C enables high-performance computing with twofolds. C++ interface enables coding as easily as ordinary floating-point numbers. My goal is convincing you to try twofolds; I think assuring accuracy of math computations is worth its cost.

Contents

## Motivation

Hereon I discuss floating-point numbers with built-in errors estimate, motivated similarly to [3]. This section is for setting-up right expectations. Expecting more or fewer leads to different techniques, like interval calculus for proving correctness, or increasing precision in hope for a better accuracy. Here I seek for compromise, assuring instead of proving results, and controlling instead of mitigating errors.

Twofold arithmetic I propose in [1] bases on Dekker 1971 formulas [5] for doubling precision. A twofold represents a real value $x$ as unevaluated sum of floating-point numbers $x_0 + x_1$, where $x_0$ approximates $x$, and $x_1$ intends estimating its deviation $\Delta x_0 = x - x_0$. Effectively twofolds double-check floating-point results via recalculating everything with nearly 2x-higher precision.

Obvious criticism is if twofolds spend computer resources reasonably. Once we have 2x-precise result, why not using it for doubling precision? Two answers are performance and control. Twofold arithmetic is much (25x times) faster than standard binary128 according to my testing. Accuracy is the cost for high performance: twofolds are accurate enough for detecting but not for compensating rounding errors.

Other similar techniques like double-double [6] might be also quite fast and give stricter results than twofolds. But I think controlling accuracy is sometimes more important than improving it. Even if I use quad precision, how can I know it is enough? In rare specific cases, I can formally prove an algorithm. But generally, proving is too difficult or does not work at all, too much depends on specific input data.

Interval calculus automates proving, but still requires special algorithms and is not always applicable. Twofolds are applicable always, and cost for that is reliability. If standard and 2x-precise results differ, result is certainly wrong, but results coinciding proves nothing. So twofolds rather test and assure than prove, hopefully finding majority of bugs caused by rounding errors, but certainly not all such bugs.

Assuring with twofolds might be static, instrumenting code with twofolds for offline debugging. Though ideally verifying should be dynamic, deal with production code, tracking rounding errors permanently in on-fly manner. So performance is critical for twofolds. Here I discuss high-performance computing tools for twofolds, plain C and C++ interfaces and experimental implementation.

C++ interface allows easily instrumenting numeric codes written in C++ by "mindlessly" [4] replacing some or all of floating-point numbers with twofolds. Overloading standard operators and functions like x+y, x<y, sqrt() et al enables programming in habitual manner. Plain C interface designed for performing at up to 100% of CPU peak without coding in assembler, though with manual vectoring for SIMD.

Of course, for twofolds 100% of peak means up to 10x times slower than ordinary floating-point, and it is question if such cost is reasonable. In [1] I discuss situations when I think it is. Modern processors are much faster than computer memory, and this gap tends to grow. So many applications utilize only part, maybe 10% of processor potential. This allows lots of underused CPU capacity that we can leverage.

My goal is convincing you to try twofolds. I think twofolds is useful tool, and improving math software reliability is worth its cost. My strategy is standardizing interface, so everyone could propose compatible implementation. My experimental implementation intends setting-up expectations. My code is free for academic and non-commercial. (For commercial, I am afraid current code version is not good enough.)

Among modern languages, I found C and C++ most appropriate for my goals. I target Windows and Linux with Microsoft and GNU compilers as platforms used by majority. Test lab was my laptop, HP Pavilion 15 with Intel Core i5-4200U processor (Haswell) 1.6 GHz (up to 2.6 GHz turbo) and Windows 8.1 (64-bits). Compiler versions: Red Hat Cygwin GCC 4.8.3, and Microsoft Visual Studio 2013 Express.

## Use examples

- Dot-product
- $Ax = f$ solver
- Corner cases
  - Summation
  - Equation root
  - Polynomial
- C++ ease

Let me start with observing brief list of twofold use examples to outline main benefits and limitations.

Dot-product example shows coding in plain C may allow very high performance, sometimes comparable to ordinary floating-point. If accessing memory is bottleneck, there remain enough underused processor capacity that twofolds can efficiently leverage.

Linear $Ax = f$ solver example shows detecting accumulation of rounding errors in ill-behaving sample systems. Important note is that twofolds can detect but generally do not resolve accuracy limitations. If standard 64-bit double precision appear not enough, probably right solution is 128-bit quads.

Corner cases example demonstrates detecting large errors in a few simple situations: solving quadratic equation, evaluating complicated polynomial, and abnormal accumulation of round-offs in summation.

C++ ease examples demonstrate programming style, almost 100% identical to ordinary floating-point.

Examples code and testing logs available at my web site [2], free for academic and non-commercial.

## Dot-product

In this subsection, I consider performance of array summation $s = \sum x_n$ and of dot-product $s = \sum x_n y_n$. Such summation is one of main sources of rounding errors in computational linear algebra, worth using higher-precision for accumulator $s$, maybe standard 128-bit quad, or double-double like XBLAS [7].

Generally, I cannot recommend twofolds as surrogate 2x-higher precision, but in this specific case, I can. Adding an ordinary number $x_i$ to 2x-precise twofold accumulator $s$ is strict operation, behaving similarly to double-double. For summation or dot-product you can choose: use twofolds for mitigating inaccuracy or only for identifying/measuring it.

Note that modern implementations of standard 128-bit quad-precision may be disappointingly slow. See my testing results in the following table. I tested with gcc of Red Hat's Cygwin 4.8.3 with my laptop built on Intel Core i5-4200U (Haswell) processor. CPU performed at 2.55 GHz in this test. Results per one CPU core, measured in millions floating-point operations per second (MFLOPS):

Table 1: GNU gcc performance (MFLOPS)

|           | add     | mul     | div     |
|-----------|---------|---------|---------|
| __float128 | 47.7532 | 36.6689 | 14.5799 |

Compare with performance for twofolds and ordinary floating-point numbers tested on same machine:

Table 2: Twofold and ordinary summation and dot-product with GNU gcc (MFLOPS)

|      | small float | large float | small doube | large double |
|------|-------------|-------------|-------------|--------------|
| tsum | 2488.63     | 2308.55     | 1247.17     | 1152.45      |
| tdot | 2440.72     | 1586.6      | 1232.54     | 793.808      |
| sum  | 14044.5     | 3548.43     | 7416.93     | 1776.22      |
| dot  | 11088.4 | 1814.15 | 5820.62 | 907.403 |

Columns are for small and large arrays of `float` and `double` type. Small array is 1K bytes and fits in CPU fast cache; large array is 64M bytes and does not fit processor cache. Twofold summation of small array does not stall on memory reading and performs around 25x times faster than `__float128`, though still about 6x times slower than summation with ordinary `double` numbers.

Losing 6x times against ordinary numbers is not good, but look at the data for large arrays. Reading data from memory is the bottleneck in this case. With ordinary numbers, this test can gain only 25% of CPU capacity in summation and only 15% in dot-product. Twofolds would leverage remaining 75-85% for the useful additional job, assessing accumulation of rounding errors, or for effectively doubling precision.

For large arrays, twofolds lose "only" 15-35% of performance to ordinary numbers. I think such cost is reasonable if you know or suspect standard double precision maybe not enough for your computation. For computational linear algebra, I guess you would suspect this often if not always. And I think in many cases even the 6x-times slow-down would be reasonable cost for assuring more reliable results.

Well, I have the above data with code manually vectorized for SIMD, specifically for Intel AVX. If relying on compiler's automatic optimization, results are much worse. The compilers I tested cannot recognize twofold operations as a good pattern for vectorization; see the following data for the GNU gcc compiler. Ordinary numbers are still quite good, but performance of twofolds looks disappointing here:

Table 3: Twofold and ordinary sum and dot-product (MFLOPS), plain C, auto-optimized with gcc

|      | small   | large   | small   | large   |
|------|--------:|--------:|--------:|--------:|
|      | float   | float   | doube   | double  |
| tsum | 364.975 | 356.466 | 358.748 | 347.326 |
| tdot | 301.3   | 285.424 | 284.726 | 253.523 |
| sum  | 6660.52 | 3620.9  | 3840.25 | 1806.46 |
| dot  | 4009.46 | 1809.63 | 2181.05 | 902.106 |

Twofold C++ interface does not support manual vectoring, and compiler automatic optimization is even less effective here. Performance depends on compiler options; following is best result I got with default options for this test with GNU compiler. While this looks 4-5x times faster than `__float128`, I of course cannot honestly recommend this for high-performance computing.

Table 4: Twofold sum and dot-product (MFLOPS), C++, auto-optimized with g++

|      | float   | float   | doube   | double  |
|------|--------:|--------:|--------:|--------:|
| tsum | 255.034 | 250.363 | 233.457 | 232.721 |
| tdot | 231.413 | 222.175 | 219.902 | 194.569 |

Vectoring of twofold operations is easy once compiler recognizes its pattern. I believe future compiler versions may support twofolds very well if we standardize C and C++ interfaces. Meanwhile, I suggest vectoring manually with plain C for high performance, or leveraging ease of C++ non-vectored twofolds for researching algorithms accuracy or debugging or if performance is not critical.

You may find these tests code, make file, and testing logs at my web site [2]. Download the file code.zip, see under the `code/xblas` folder.

## $Ax = f$ solver

Solving linear $Ax = f$ system is probably central problem for numeric computations as many algorithms include this step. In this subsection I consider a simple LU solver (with pivoting by rows) for general-case square matrix $A$. This example demonstrates detecting accumulation of rounding errors with well- and ill-conditioned matrix $A$. Secondarily, this example demonstrates ease of coding with twofolds in C++.

Please download the example sources and testing logs from my web site [2], archive code.zip, see folder code/lups. Type-generic function `lups_impl()` found in the source file `lups.cpp` implements this solver for all types, including ordinary floating-point `float` and `double` and `__float128`, as well as for the generic types `twofold<T>` and `coupled<T>` where a type T may be `float` or `double`.

Obviously, type `twofold<T>` implements a twofold, which is pair of "value" and "error" variables of the basic type T. In the calculations, the value part behaves 100% same way as ordinary numbers of type T, and error part estimates the accumulation of rounding errors. This must allow detecting if accumulated error appears too large. Or you can consider value+error which is often more accurate than value alone.

Auxiliary type `coupled<T>` is special case of twofold with value and error parts "renormalized", that is error part small comparing value so mantissa bits of value and error do not overlap. This corresponds to Dekker [5] approach to higher precision "coupled-length" arithmetic. Such "coupled" arithmetic often provides results nearly accurate as standard quad or double-double [6], but generally is not as strict.

Well known that linear system may solve badly even if small dimension. I test this solver against 5x5 systems with a Jordan cell as matrix $A$, or against such system perturbed by exchanging its rows:

$$\begin{pmatrix} \lambda & 1 & \\ & \ddots & 1 \\ & & \lambda \end{pmatrix} \begin{pmatrix} x_1 \\ \vdots \\ x_n \end{pmatrix} = \begin{pmatrix} f_1 \\ \vdots \\ f_n \end{pmatrix}$$

Numeric solution would behave badly if $\lambda$ is small. I test with $\lambda = 10^{-4}$, which causes $x_{n-1}$ be 4 decimal orders less accurate than $x_n$, etc. For single precision (type `float`), $x_1$ would be completely wrong, and for `double`, $x_1$ would be inaccurate. The testing must show if twofolds can identify this inaccuracy, and ideally, measure it more-or-less adequately.

My testing log file `luptest.gcc.log` contains numeric solutions with ordinary `float`, `double`, and quad-precision (`__float128`) numbers, and with twofold and coupled over `float` and `double` types.

If not rounding errors, solution must be exact. To check that, I test variant of system multiplied by $10^4$ so all coefficients are integer. Another test is "normalized" so numeric $\lambda \approx 10^{-4}$ is a bit inexact due to rounding error. Then I test "normalized and truncated" variant representing $\lambda$ as twofold with nullified error part. For testing with quad precision, precision of was $\lambda$ truncated to `double`.

Consider the following testing results for "normalized" but not "truncated" twofolds over `double`:

```
lupstest: twofold<double>
problem: ill_10k, normalized
x (result):
    1.11012 [-0.110123   ]
    0.999989[ 1.10123e-05]
    1       [-1.10123e-09]
    1       [ 1.10134e-13]
    1       [0]
```

Expected resulting $x$ must be all unities exactly. Computed result is inexact, and error part of twofolds (numbers in square brackets) correctly identify this problem, moreover adequately measure rounding error accumulated in $x$. Measuring the error such adequately is possible because twofold value + error combination is accurate as nearly 2x-precise, and such nearly-2x precision is enough for this example.

If we truncate precision of $\lambda$, twofold seems to underestimate accumulated error (but really it does not):

```
lupstest: twofold<double>
```

```
problem: ill_10k, normalized, truncated
    x (result):
    1.11012  [ 4.79169e-05]
    0.999989 [-4.79169e-09]
    1        [ 4.79169e-13]
    1        [-4.79217e-17]
    1        [0]
```

The solver does not "know" how much we truncated accuracy of matrix $A$, and assumes the incoming inaccuracy is zero. With such assumptions, $x_1 \approx 1.11012$ is correct solution, to check this please see quad-precision results below in the testing log. Moreover, because double-double precision is enough for this example, assessing rounding errors accumulated in $x_1$ like $4.79169 \cdot 10^{-5}$ is adequate.

Now let us consider solving same system with twofold over `float`, without truncating $\lambda$'s precision:

```
lupstest: twofold<float>
problem: ill_10k, normalized
    x (result):
    -1.65923e+08 [ 1.65923e+08]
    16593.3      [-16592.3]
    -0.659227    [ 1.65923]
    1.00017      [-0.000165939]
    1[0]
```

Single precision is certainly not enough for this example, so solution ruins completely. Even 2x-higher precision of twofold is worse than `double` alone thus is not enough as well. So twofold might be not accurate in measuring rounding errors, but looks good in identifying the fact of error getting too large. For example, for $x_3$ we see that its value ($\approx -0.659227$) is less than its error estimate ($\approx 1.65923$).

Indeed, we could evaluate value + error ($-0.659227 + 1.65923 = 1.000003$) as good approximation for $x_3$. But for $x_2$ this trick would gain only two significant digits, and no significant digits at all for $x_1$. Thus I think, more reasonable would be concluding that `float` precision is not enough for solving this example, that this case probably requires at least `double` or maybe even higher precision.

Finally, look at twofolds over `float` if we truncate error parts of matrix $A$ coefficients:

```
lupstest: twofold<float>
problem: ill_10k, normalized, truncated
    x (result):
    -1.65923e+08 [-25280.1]
     16593.3     [ 2.52766]
    -0.659227    [-0.00025268]
     1.00017     [ 2.52663e-08]
     1[0]
```

Solution is same wrong for $x_1$ and $x_2$ due to lack of the `float` precision, and twice-as-`float` precision of twofold is not enough as well. In such situation, twofolds of course cannot compensate the error, and even fail detecting the fact of the problem. This is not good, but not too bad actually. Let me explain:

Unlike intervals, twofolds do not tend over-estimating accuracy problems. Informally, if standard single or double precision were enough, twofolds would just confirm the standard result. If standard precision is not enough, twofolds detect and even measure the error if 2x-precise approximation allows. In worst case, if 2x-precision is not enough, twofolds may still detect the problem or may occasionally miss it.

Such behavior perfectly fits ideology of assuring by testing: a good test must not panic in vain. Question is if testing with twofolds is worth resources: if it can detect majority of problems and improve numeric software reliability significantly. I think answering honestly requires further experimentation.

## Corner cases

- [Summation](#)
- [Equation root](#)
- [Polynomial](#)

In this sub-section, let us consider examples of severe rounding errors in simple math computations; solving quadratic equation, evaluating a polynomial, and accumulating rounding errors in summation. You may find these example sources and test logs at my web site [2], see under `code/corner_cases` folder in the code.zip archive.

### Summation

This well-known example already considered in [1]. It demonstrates how suddenly accumulation of round-offs may grow. Normally, rounding error tend to compensate mutually, so accumulated error grows moderately in typical computations. However sometimes, rounding errors accumulate much faster than typically, which may lead to catastrophic consequences if computation is mission-critical.

This is example of such non-typical rounding errors accumulation that caused real techno-genic accident with severe consequences. The problem happened with a device with timer assumed to count seconds so increasing by 1/10 ten times per second. The counter variable was floating-point of single precision. At the interval of 100 hours, which is 3 600 000 ticks, accumulated error would be around 3½ hours.

Look how such counter might work if implemented with twofolds or intervals:

```
test: type=twofold<float>, hours=100
    1/10 s: 0.1[-1.49012e-09]
    result: 96.3958[3.54008] hours
    expect: 100 hours

test: type=interval<float>, hours=100
    1/10 s: [0.1,0.1]
    result: [94.6528,111.328] hours
    expect: 100 hours
```

Interval guarantees boundary which result would fit even in worst case if rounding unfortunate in every of 3 600 000 summations. Unlike that, twofolds assess rounding as they were in reality specifically in this computation. This is fundamental difference: analyzing specific case must be easier usually. The cost for such ease is that twofold may accumulate error itself, like here value + error deviates from correct result by around 0.1% (that is: 96.3958 + 3.54008 = 99.93588, which differs by ~0.64% from 100 hours).

### Equation root

This example also considered in [1] is solving quadratic equation $ax^2 + bx + c = 0$ with school formula:

$$x_{0,1} = \frac{-b \pm \sqrt{b^2 - 4ac}}{2a}$$

Let $a = 1$ and $b = 2$, and try $c$ very close to 0 or to 1, specifically $c = 10^{-8}$ or $c = 1 \pm 10^{-8}$. First case examines accuracy loss in $-b + \sqrt{b^2 - 4ac}$ with $d = \sqrt{b^2 - 4ac}$ very close to $b$, and second case tests square root of inaccurate argument very close to zero which may accidentally result in NaN.

Note that binary32 type (`float` of C/C++) cannot represent $1 \pm 10^{-8}$, thus $b^2 - 4ac$ would be wrong with single precision. For `double` type, this formula must cause losing around half of significant digits.

Following is fragment from my test log:

```
test: type=float
  parameters:
    a: 1[0]
    b: 2[0]
    c: 1e-08[6.07747e-17]
    d: 2[1e-08]
    x0: -2[-5e-09]
    x1: 0[5e-09]

test: type=double
  parameters:
    a: 1[0]
    b: 2[0]
    c: 1e-08[0]
    d: 1.99999999[3.57747092909712e-17]
    x0: -1.999999995[-1.28909657108001e-16]
    x1: -5.00000008063495e-09[1.78873546454856e-17]
```

Here $c = 10^{-8}$, so root of discriminant $d = \sqrt{b^2 - 4ac}$ must equal $\sqrt{3.99999996} \approx 2 - 10^{-8}$, and result must be $x_0 \approx -1.999999995$ and $x_1 \approx -5.0000000125 \cdot 10^{-9}$.

Double precision fits these expectations, and accuracy estimate for $x_1$ predictably says around half of significant digits is lost. Float precision is not enough, and twofold correctly assesses result inaccuracy; particularly, $x_1$ is completely wrong in `float` case.

Then, note that twofold square root function correctly processes a negative argument:

```
test: type=float
  parameters:
    a: 1[0]
    b: 2[0]
    c: 1[1e-08]
    d: 0[nan]
    x0: -1[nan]
    x1: -1[nan]

test: type=double
  parameters:
    a: 1[0]
    b: 2[0]
    c: 1.00000001[0]
    d: nan[nan]
    x0: nan[nan]
    x1: nan[nan]
```

Here $c = 1 + 10^{-8}$, thus discriminant $b^2 - 4ac$ must be negative so $d$ must be NaN. For `double` type it is. For `float` type, value part of $c$ is exactly 1, so value parts of $d$, $x_0$, $x_1$ equal to 0 and $-1$ accordingly. However, twofold value + error is able to detect the problem in this example, so error parts of solution correctly assigned to NaN.

### Polynomial

Let us consider evaluating a polynomial, specifically the one by S. Rump [8]. This example remarkable behavior is that single, double, and extended precision confirm same result, which is wrong however. Here is Rump polynomial with $a = 77617$ and $b = 33096$, computing literally like printed here:

$$f = 21 \cdot b \cdot b - 2 \cdot a \cdot a + 55 \cdot b \cdot b \cdot b \cdot b - 10 \cdot a \cdot a \cdot b \cdot b + a/2b$$

Results according to Rump and as reproduced in my testing is following. Here extended precision is Intel 80-bits floating available as `long double` and "quad" is `__float128` as tested with GNU compiler:

single     precision $f \approx 1.172603\cdots$
double     precision $f \approx 1.1726039400531\cdots$
extended precision $f \approx 1.172603940053178\cdots$
quad       precision $f \approx -0.827386\cdots$

Only the last (quad) result is correct and equals to expected $f = a/2b - 2 \approx -0.827386$, while others are wrong due to higher-order terms of the polynomial cancelling to 0 instead of correct $-2$. Following are twofold results:

twofold over `float` , $f \approx 1.1726\,[\,-2.47524 \cdot 10^{-8}\,]$
twofold over `double`, $f \approx 1.1726\,[\,-2\,]$

As we could expect, twice-as-single precision is not enough for correctly assessing the inaccuracy, so twofold over `float` erroneously confirm the wrong result. In turn, twice-as-double precision allows twofold over `double` to assess the inaccuracy correctly in this example.

Now let us play a little bit more with this Rump example. This polynomial is very sensitive to rounding errors, so reordering calculations may change result completely if precision is not enough. Following is result of computing differently, like $f = (21b^2 - 2a^2) + (55b^4 - 10a^2b^2) + a/2b$, with twofolds:

twofold over `float` , $f \approx -4.38709 \cdot 10^{12}\,[\,4.38709 \cdot 10^{12}\,]$
twofold over `double`, $f \approx 2687.17\,[\,-2688\,]$

Again, twice-as-double precision allows twofold over `double` to identify and measure the error. More remarkable is behavior of twofold over `float`. Its twice-as-single precision must not allow measuring or even identifying the problem, but it identifies and somewhat measures though very inaccurately. This is important fact about twofolds: they do much better in detecting accuracy problems than expected.

Identifying problems is fundamentally easier than resolving. If calculation is very sensitive to precision, twofold recalculating with 2x- precision may identify this sensitivity, even if precision is not enough for computing the correct result. I think this must help finding many bugs caused by lack of precision.

### C++ ease

This subsection briefly overviews a few simple use examples you may find at my web page [2], archive code.zip, under `code/examples` folder. This folder contains several files names like `ex_*.cpp` which demonstrate using twofolds with C++.

They demonstrate the generic types for twofold and coupled-length arithmetic, creating and initializing variables of such types, and arithmetic and logical operations over such variables. Following short code fragment demonstrates printing:

```
double pi = 3.141592653589793;
double  e = 2.718281828459045;
coupled<T> p = e;
twofold<T> t = pi;
cout << "    p : " << p << endl; // same as cout << to_string(p)
cout << "    t : " << t << endl;
```

In this fragment, parameter type T may be `float` or `double`. Operating with twofolds is same easy as ordinary numbers; you can mindlessly replace all `float` variables with `twofold<float>` and similarly for `double`. Arithmetic over `coupled<T>` requires additional care. For details, see the examples code.

Unlike intervals, if you mindlessly replace all or some of floating-point variables with twofolds this must not change your program behavior. If compiler does not reorder operations, twofold value parts would do 100% same as corresponding variables in original program, modulo additional memory for error parts and computing slower. If you ignore error parts, results must be bitwise same as original.

This way you can assess accuracy of computations without risk of damaging results. In worst case, twofolds may fail finding accuracy problems. In best case, you can find majority of accuracy bugs.

# Twofold arithmetic

- [Algorithms](#)
- [Fast vs strict](#)
- [Performance](#)
- [Perspectives](#)

## Algorithms

In this article, I propose C and C++ interfaces for the "Twofold fast arithmetic" I suggested in [1]. In this section, I would briefly remind important points about twofolds, what they are and what they are not, and discuss the performance expectations. For the twofold arithmetic algorithms, please see [1].

## Fast vs strict

As said in Use examples/Dot-product section above, twofold fast arithmetic over `double` is around 25x times faster than standard 128-bit quad according to my testing. High performance has cost; twofolds arithmetic is not strict. Twofold error part may be not accurate; it may lose a few bits of significance.

Trading strictness for performance is purposeful design decision. Twofolds use fastest formulas I know with minimally acceptable accuracy; right enough for detecting but not for resolving accuracy problems. Detecting errors is fundamentally easier than resolving; so twofolds arithmetic designed for detecting.

I am writing this to emphasize, please do not consider twofolds as a surrogate for faster quad precision. If standard `double` is not enough, consider `__float128` or `_Quad` types of GNU and Intel compilers, or maybe multi-precision pack like GNU MPFR (http://mpfr.org/) or similar, maybe double-double et al [6].

The only but important exception: twofold arithmetic is strict in vector summation and dot-product operations like described in Use examples/Dot-product above. You may use twofolds for increasing precision to nearly-twice-as-`double` in this specific case. See also XBLAS [7] about this case.

## Performance

Performance is key point for twofolds; let us consider more details in this subsection. Here I display my testing results and explain them. The tests code, build-and-run scripts, and testing logs available at my web site [2], download code.zip archive, see under `code/perftest` folder.

My testing lab is my laptop, HP Pavilion 15 built on Intel Core i5-4200U (Haswell) processor with flexible frequency, which worked at 2.25 to 2.55 GHz in my testing. OS is Windows 8.1 (64 bits). C/C++ compiler versions, both 64-bits variant: Red Hat Cygwin 4.8.3, and Microsoft Visual Studio 2013 Express, available free at the vendor web sites. Here I discuss my results with the GNU compiler.

The test iterates the twofold add, multiply, divide, and square root operations and prints the measured performance in mega-operations-per-second (MOPS). The input and output data organized into arrays of 1K (=1024) bytes. Arrays allow vectoring operations with Intel AVX for higher performance, while not suffering slow memory read/write as 1K is short enough to fit into processor's fast cache.

By default, my twofolds implementation uses Intel AVX, and additionally supports manual vectoring if compiled with –DAVX option. Here I discuss default (scalar) and vectored results. Default performance appears poor, but manual vectoring improves it around 4x times for `double` and 8x times for `float`.

Following is default (scalar) data for plain C and for C++. Processor frequency was 2.55 GHz as I could observe with Windows Task Manager:

Table 5: C++ twofold performance (MOPS)

|        | tadd    | tmul   | tdiv    | tsqrt   |
|--------|---------|--------|---------|---------|
| double | 299.809 | 307.45 | 91.3929 | 60.6951 |

|        | tadd    | tmul    | tdiv    | tsqrt   |
|--------|---------|---------|---------|---------|
| float  | 317.945 | 316.957 | 181.048 | 110.134 |

Table 6: Plain C twofold performance (MOPS)

|        | tadd    | tmul    | tdiv    | tsqrt   |
|--------|---------|---------|---------|---------|
| double | 304.078 | 297.1   | 91.4521 | 60.7901 |
| float  | 318     | 318.262 | 181.473 | 106.052 |

Float tadd() performance at 318 MOPS looks poor, but is nearly best we should expect. Recall from [1] that twofold-add requires 8 basic add/subtract operations; so 318 million twofold adds is 2544 million basic operations. Doing 1 operation per CPU tick, this corresponds to 2.544 GHz, while CPU performed at 2.55 GHz as I observed. So twofold-add gains up to 99% of CPU theoretical peak.

Following are plain C results with manual vectoring for AVX. Table lines 256_pd and 256_ps correspond to vectoring `double` and `float` types with AVX 256-bit registers, 128_pd and 128_ps are for half-length 128-bit registers, and 128_sd and 128_ss use 128-bit AVX registers for scalar `double` and `float` types.

Processor operated at 2.55 GHz while testing scalar operations, periodically dropping to 2.25 GHz while doing vectored operations with 128- and 256-bit registers. Obviously, these frequency drops were due to higher load so heating of CPU with vector operations, which caused automatic frequency throttling:

Table 7: Plain C twofold performance (MOPS)

|        | tadd    | tmul    | tdiv    | tsqrt   |
|--------|---------|---------|---------|---------|
| 256_pd | 1144.34 | 1610.73 | 162.739 | 108.928 |
| 256_ps | 2288.67 | 3225.43 | 644.849 | 420.525 |
| 128_pd | 633.018 | 909.824 | 162.546 | 108.328 |
| 128_ps | 1264.56 | 1818.02 | 638.28  | 431.31  |
| 128_sd | 316.17  | 482.211 | 91.4511 | 60.9699 |
| 128_ss | 315.781 | 484.1   | 181.496 | 120.543 |

As expected, vectoring for single-instruction-multi-data (SIMD) speeds-up computations linearly. Using 128-bit AVX registers increases performance around 2x times for `double` and 4x for `float`, and 256-bit registers increases 4x and 8x times accordingly.

Performance like 1144 MOPS in twofold-add with 256-bit registers of `double` data (line: 256_pd) is 9152 (=1144*8) gigaflops of basic add/subtracts, which is 102% of this processor theoretical peak at 2.25 GHz (so I think frequency was somewhat higher in average). Performance as 316 MOPS in twofold-add with 128_pd is 2528 megaflops of basic add/subtracts or 99% of best possible for this CPU at 2.55 GHz.

This must prove that implementation and test encoded quite well without obvious performance gaps. Even so, twofolds are several times (up to 8x) slower than ordinary `double` and `float` numbers. I think future processors and compilers can mitigate this gap if twofolds accepted widely enough by industry.

### Perspectives

I think standardizing C and C++ interfaces for twofolds must allow future compiler versions to support automatic vectoring of twofold computations for higher performance. Technically, vectoring twofolds is trivial once compiler recognizes this pattern. Twofold functions must make recognizing very easy.

In [1], I also discuss possible support in hardware, so future processors could perform twofolds faster. Ideally, twofold sum might be only 3x times slower than regular numbers, and only 2x times in special important case of array summation or dot-product.

# C/C++ interface

- 

## Open standard

In this section, I briefly outline main points about C and C++ interfaces for twofold arithmetic. The idea is standardizing the interfaces, so that anyone could provide compatible implementation. Ideally, twofolds interface should become part of C and C++ language standard. Additionally we need non-standard processor-specific extensions like AVX intrinsic for manual optimization. See AVX intrinsic guide [9].

## C/C++ keywords

If twofolds accepted widely enough, I think plain C should support it with keywords like `_Twofold` and `_Coupled`, similarly to `_Complex` in C99 and later versions of the plain C language:

```
double _Twofold x; // twofold
float  _Coupled y; // coupled-precision (special re-normalized twofold)
```

Important details like constructing and decomposing twofolds in plain C and combining `_Twofold` and `_Coupled` keywords are out of this article's scope.

Above that, I would not propose any structured types for twofold and coupled-precision numbers in plain C. For twofolds arithmetic, I propose low-level functions like following. Note the suffix 'f' in the function name if `float` type:

```
double x0,x1, y0,y1, z0,z1;
z0 = tadd(x0,x1,y0,y1,&z1); // twofolds: (z0+z1) := (x0+x1) + (y0+y1)

float u0,u1, v0,v1, w0,w1;
w0 = paddf(u0,v1,v0,v1,&w1); // coupled: (w0+w1) := (u0+u1) + (v0+v1)
```

Such low-level functions must be easy for compiler optimization, allow maximal use of CPU registers. Then I propose non-standard expanding this low-level interface for maximal use of Intel AVX registers. Similar hardware specific expansions must work for other SIMD processors:

```
__m256d x0,x1, y0,y1, z0,z1;           // AVX 256-bit registers as double
z0 = _mm256_tadd_pd(x0,x1,y0,y1,&z1); // (z0+z1) := (x0+x1) + (y0+y1)
```

C++ interface overloads functions `tadd()` et al for `float` arguments without a type-specific suffix 'f' in function name:

```
float u0,u1, v0,v1, w0,w1;
```

```
w0 = padd(u0,v1,v0,v1,&w1); // coupled: (w0+w1) := (u0+u1) + (v0+v1)
```

Note that I do not propose such overloading for non-standard AVX types __m256 and __m256d. Generally, I do not propose C++ interface supporting processor-specific extensions. Supporting processor specifics in plain C is enough in my view.

For C++, I propose generic types `twofold<T>` and `coupled<T>`, where type T may be `float`, `double`, or `long double`. (Current experimental implementation does not supports `long double` however.)

```
twofold<double> x; // twofold
coupled<float>  y; // coupled-precision (special re-normalized twofold)
```

Arithmetic functions like `tadd()` would accept arguments of twofold type and produce twofold results:

```
twofold<T> x, y, z; // twofolds
z = tadd(x,y);      // z = x + y

coupled<T> u, v, w; // coupled-precision
w = padd(u,v);      // w = u + v
```

Convenience constructors and operators allow operating twofolds same way like ordinary numbers:

```
twofold<double> x, y, z;
x = 3.141592653589793;
y = 2.718281828459045;
z = x - y;
if (z > 0) ...
cout << "pi – e = " << z << endl;
```

Convenience operators available only for twofolds, and do not support coupled-precision arithmetic. If you need it, please use functional interface, like `padd(u,v)`, or explicitly renormalize twofold results. You may use fast-renormalizing if you are sure error part of result is small comparing value:

```
coupled<T> u, v, w;       // coupled-precision
w = fast_renorm(u + v);   // w = u + v
```

Please note anyway, that coupled-precision arithmetic is auxiliary, and might be not strict enough for increasing precision. Probably, standard quad would be more appropriate if you need better accuracy.

### Plain C interface

Plain C interface is low-level. It does not introduce any types for twofold and coupled-precision data, but defines set of functions over standard `float` and `double` types, treating twofold as pair of parameters.

Yet I do not support `long double` type, as twofold arithmetic over extended-precision would not be fast enough on modern processors. Specifically, twofold arithmetic needs fast fused-multiply-add (FMA), which AMD and Intel processors I target do not support for extended-precision floating-point numbers.

Returning pair of numbers forming resulting twofold is somewhat tricky: function returns main (value) part of twofold normally, and reserves extra argument for pointer to auxiliary (error) part of resulting twofold. For example, summation of twofold $x_0 + x_1$ with $y_0 + y_1$ resulting in $z_0 + z_1$:

```
double x0,x1, y0,y1, z0,z1; // twofolds: (x0+x1), (y0+y1), (z0+z1)
z0 = tadd(x0,x1,y0,y1,&z1); // return z0 normally, z1 with pointer
```

Modern compilers can optimize such calls very well: pass parameters via CPU registers, and remove redundant data moving at all as inline function allows. My testing with GNU and Microsoft C and C++ shows such interface allows utilizing up to 100% of CPU capacity, with zero overhead on functions calls.

The plain C interface defines the following functions for twofold summation over double type. If we say "shaped" for twofold or coupled versus "dotted" for ordinary numbers, main function `tadd()` assumes both arguments are shaped, and additional functions assume 1$^{st}$ or 2$^{nd}$ or both arguments are dotted:

| Function | Comment |
|---|---|
| `z0 = tadd (x0,x1,y0,y1,&z1)` | Both arguments twofold |
| `z0 = tadd2(x0   ,y0,y1,&z1)` | Only 2$^{nd}$ argument twofold |
| `z0 = tadd1(x0,x1,y0   ,&z1)` | Only 1$^{st}$ argument twofold |
| `z0 = tadd0(x0   ,y0   ,&z1)` | Twofold sum of dotted arguments |

A coupled-precision function is named have prefix 'p' instead of 't'. Such p-functions assume but do not check if input argument renormalized so error part is small comparing value. Ensuring this pre-condition is programmer's responsibility. Note that `tadd0()` and `padd0()` compute the same:

| Function | Comment |
|---|---|
| `z0 = padd (x0,x1,y0,y1,&z1)` | Both arguments coupled-precision |
| `z0 = padd2(x0   ,y0,y1,&z1)` | Only 2$^{nd}$ argument coupled-precision |
| `z0 = padd1(x0,x1,y0   ,&z1)` | Only 1$^{st}$ argument coupled-precision |
| `z0 = padd0(x0   ,y0   ,&z1)` | Coupled sum of dotted arguments |

Same functions for float type named with suffix 'f':

| Twofold | Coupled |
|---|---|
| `z0 = taddf (x0,x1,y0,y1,&z1)` | `z0 = paddf (x0,x1,y0,y1,&z1)` |
| `z0 = tadd2f(x0   ,y0,y1,&z1)` | `z0 = padd2f(x0   ,y0,y1,&z1)` |
| `z0 = tadd1f(x0,x1,y0   ,&z1)` | `z0 = padd1f(x0,x1,y0   ,&z1)` |
| `z0 = tadd0f(x0   ,y0   ,&z1)` | `z0 = padd0f(x0   ,y0   ,&z1)` |

Similarly are defined groups of functions for operations of subtracting, multiplying, dividing, and taking square root. Except for square root, there is no need in variants like sqrt2 and sqrt1. Enlisting explicitly, where {t|p} means selecting a mandatory prefix, and selections in square brackets are optional:

| Function name | Description |
|---|---|
| `{t|p} add [0|1|2] [f]` | Add |
| `{t|p} sub [0|1|2] [f]` | Subtract |
| `{t|p} mul [0|1|2] [f]` | Multiply |
| `{t|p} div [0|1|2] [f]` | Divide |
| `{t|p} sqrt [0] [f]` | Square root |

Three additional functions for multiplying, dividing, and square root, compute twofold result faster if arguments are coupled-precision, using the special property that error is much smaller than value for coupled numbers. These function named with suffix 'p'. Enlisting explicitly:

| Function name | Description |
|---|---|
| `tmulp [f]` | Multiply |
| `tdivp [f]` | Divide |
| `tsqrtp [f]` | Square root |

Two special functions renormalize twofold into coupled-precision numbers. Function `renormalize()` is for general case, and `fast_renorm()` is applicable in case if argument is almost re-normal already, that is argument's error part is small versus value. Granting this pre-condition is programmer's responsibility.

Auxiliary function `fast_add0()` implements fast-renormalizing, and `fast_sub0()` accompanies it. Here is overall list of the additional functions:

| Over double | Over float |
|---|---|
| z0 = renormalize(x0,x1,&z1) | z0 = renormalizef(x0,x1,&z1) |
| z0 = fast_renorm(x0,x1,&z1) | z0 = fast_renormf(x0,x1,&z1) |
| z0 = fast_add0(x0,x1,&z1) | z0 = fast_add0f(x0,x1,&z1) |
| z0 = fast_sub0(x0,x1,&z1) | z0 = fast_sub0f(x0,x1,&z1) |

To conclude, following is the full list of twofold and coupled-precision functions for plain C:

- Main list: {t|p}{add|sub|mul|div}[0|1|2][f] and {t|p}sqrt[0][f]
- Special list: t{mulp|divp|sqrtp}[f]
- Auxiliary: renormalize[f], fast_renorm[f], fast_add0[f], fast_sub0[f]

## AVX extensions

Performance is critical for twofolds, and optimizing for specific processor is necessary. Twofold interface would consist of two parts: standard, and processor-specific. Standard part deals with C/C++ standard floating-point types, and non-standard would support manually vectoring for SIMD, like AMD/Intel AVX.

Modern processors by Intel and AMD support so-called AVX extensions, 256-bit registers and assembler instructions for "packed" operations over 4 of `double` or 8 of `float` values at once. GNU and Microsoft compilers support manual optimizing for AVX in C and in C++ with AVX "intrinsics", processor-specific vector types `__m256` and `__m256d` and packed functions that map directly to assembler instructions:

```
__mm256d x, y, z;        // x, y, z are vectors, each is 4 of double
z = _mm256_add_pd(x,y); // packed summation: z[i]=x[i]+y[i], i=0,1,2,3
```

My proposal is supporting AVX with twofold operations, like for example twofold summation:

```
__mm256d x0,x1, y0,y1, z0,z1;              // twofold: x0[i] + x1[i]
z0 = _mm256_tadd_pd(x0,x1,y0,y1,&z1); // packed twofold summation
```

Similarly to scalar `tadd()`, packed function returns result's value part $z_0$ normally and error part $z_1$ via pointer. This allows compiler optimizing function calls, transferring parameters via processor registers, and eliminating needless data movement if an inline function. My testing with GNU and Microsoft C and C++ shows, modern compiler can gain almost 100% of processor capacity with zero overhead for calling.

Such interface allows more-or-less logically instrumenting code written with AMD/Intel AVX extensions. In the above example, the value part x0, y0, z0 would behave bitwise same way as original x, y, z, and auxiliary error part x1, y1, z1 would allow tracking rounding errors. Such tracking costs, but would not change behavior of original algorithm.

Same operation over packed `float` data has "_ps" suffix in function name:

```
__mm256 x0,x1, y0,y1, z0,z1;              // packed float: x[i], i=0,..,7
z0 = _mm256_tadd_ps(x0,x1,y0,y1,&z1); // packed twofold summation
```

Similarly to scalar `tadd()`, function variants with digit 0, 1, or 2 in its name assume that one or both of arguments are dotted (not twofold) numbers:

```
__mm256 x0,x1, y0,y1, z0,z1;
```

```
z0 = _mm256_tadd_ps (x0,x1,y0,y1,&z1); // both arguments twofold
z0 = _mm256_tadd2_ps(x0   ,y0,y1,&z1); // only 2nd argument twofold
z0 = _mm256_tadd1_ps(x0,x1,y0   ,&z1); // only 1st argument twofold
z0 = _mm256_tadd0_ps(x0   ,y0   ,&z1); // twofold sum of dotted arguments
```

Coupled-precision variant has prefix 'p' instead of 't' in the function's name. Note that `padd0()` and `tadd0()` compute the same result, but still have different names:

```
__mm256 x0,x1, y0,y1, z0,z1;
z0 = _mm256_padd_ps (x0,x1,y0,y1,&z1); // both arguments twofold
z0 = _mm256_padd2_ps(x0   ,y0,y1,&z1); // only 2nd argument twofold
z0 = _mm256_padd1_ps(x0,x1,y0   ,&z1); // only 1st argument twofold
z0 = _mm256_padd0_ps(x0   ,y0   ,&z1); // coupled sum or dotted arguments
```

Similarly named groups of functions would serve for twofold/coupled subtracting, multiplying, dividing, and taking square root. Obviously, square root does not need variants with index 1 or 2. Following is the list of such functions. Here {p|t} and {d|s} means selecting prefix and suffix, and selection like [0|1|2] in square brackets is optional:

| Function | Description |
|---|---|
| _mm256_{t|p}add[0|1|2]_p{s|d} | Add |
| _mm256_{t|p}sub[0|1|2]_p{s|d} | Subtract |
| _mm256_{t|p}mul[0|1|2]_p{s|d} | Multiply |
| _mm256_{t|p}div[0|1|2]_p{s|d} | Divide |
| _mm256_{t|p}sqrt[0]_p{s|d} | Square root |

Besides of 256-bit, AVX supports packed and scalar operations with half-length 128-bit registers. Scalar means processing only one `float` or `double` value in the lower 32 or 64 bits of 128-bit register. Intrinsic functions for such operations look like following. Please note "pd" versus "sd" suffix in function name:

```
__mm128d x, y, z;    // x, y, z are vectors, each 2 of double
z = _mm_add_pd(x,y); // packed summation: z[i]=x[i]+y[i], i=0,1
z = _mm_add_sd(x,y); // scalar summation: z[0]=x[0]+y[0], z[1] intact
```

Twofold interface would support packed and scalar operations with 128-bit registers like following:

```
__mm128d x0,x1, y0,y1, z0,z1;          // twofold: x0[i] + x1[i]
z0 = _mm_tadd_pd(x0,x1,y0,y1,&z1); // packed twofold summation
z0 = _mm_tadd_sd(x0,x1,y0,y1,&z1); // scalar twofold summation
```

Explicit list of twofold/coupled functions for add, subtract, multiply, divide, and square root operations with 128-bit types. Here additional {s|p} selection chooses between scalar and packed:

| Function | Description |
|---|---|
| _mm_{t|p}add[0|1|2]_{s|p}{s|d} | Add |
| _mm_{t|p}sub[0|1|2]_{s|p}{s|d} | Subtract |
| _mm_{t|p}mul[0|1|2]_{s|p}{s|d} | Multiply |
| _mm_{t|p}div[0|1|2]_{s|p}{s|d} | Divide |
| _mm_{t|p}sqrt[0]_{s|p}{s|d} | Square root |

Special function compute twofold multiplying, dividing, and square root faster in case if arguments are coupled-precision, using the assumption that error part of argument is very small comparing value part. Granting such pre-condition is programmer's responsibility. Scalar operations not available for 256-bit:

| 256 bits | 128 bits |
|---|---|
| `_mm256_tmulp_p{s\|d}` | `_mm_tmulp_{s\|p}{s\|d}` |
| `_mm256_tdivp_p{s\|d}` | `_mm_tdivp_{s\|p}{s\|d}` |
| `_mm256_tsqrtp_p{s\|d}` | `_mm_tsqrtp_{s\|p}{s\|d}` |

Auxiliary "renormalize" function converts twofold to coupled and "fast_renorm" converts in assumption that error part of argument is not greater (by magnitude) than its value. Function fast_add0 implements fast renormalizing, and fast_sub0 accompanies it. Interface explicitly:

| 256 bits | 128 bits |
|---|---|
| `_mm256_renormalize_p{s\|d}` | `_mm_renormalize_{s\|p}{s\|d}` |
| `_mm256_fast_renorm_p{s\|d}` | `_mm_fast_renorm_{s\|p}{s\|d}` |
| `_mm256_fast_add0_p{s\|d}` | `_mm_fast_add0_{s\|p}{s\|d}` |
| `_mm256_fast_sub0_p{s\|d}` | `_mm_fast_sub0_{s\|p}{s\|d}` |

To conclude, following is the full list of twofold and coupled-precision functions for AVX in plain C:

- Main list: add, sub, mul, div, sqrt              – e.g.: _mm256_tadd_pd
- Special list: mulp, divp, sqrtp                  – e.g.: _mm256_tmulp_pd
- Auxiliary: renormalize, fast_renorm, fast_add0, fast_sub0  – e.g.: _mm256_renormalize_pd

## C++ interface

- Types
- Functions
- Operators
- Comparing
- Namespace
- Printing

### Types

C++ interface founds on top of plain C. It does not add much functionality; it adds convenience.

Basically, for C++ we would define two generic types, structures for twofold and coupled-precision data. Here parameter type T may be standard `float` or `double` or `long double`, though my implementation targeting Intel (and AMD) processors does not support extended precision:

```
template<typename T> struct twofold { T value, error; … };
template<typename T> struct coupled: public twofold<T> { … };
```

Structure fields for value and error parts of twofold or coupled number are available publicly. This allows damaging coupled-precision invariant, that error must be small comparing value so that value and error mantissas do not overlap. This is programmer's responsibility to ensure this invariant.

Inheriting `coupled<T>` from `twofold<T>` allows assigning a coupled-precision value to twofold variable but not conversely, which corresponds to coupled-precision numbers being special case of twofolds. For converting a twofold to coupled, please use renormalize functions, constructors, or explicit type cast.

Constructors allow making twofold or coupled value from dotted value of same or different basic type T, for example:

```
twofold<double> e = 2.71828; // same basic type (double)
twofold<float>  pi = 3.14159; // make twofold<float> from double
twofold<float>  zero = 0;     // make twofold<float> from int
```

Creating from another twofold or coupled of same or different basic type may require explicit type cast:

```
twofold<double> t;
coupled<double> p;
twofold<float> tf;
coupled<float> pf;
t = p;                   // no cast required, just copy coupled to twofold
t = (twofold<double>)pf; // explicit cast: reshape to the other basic type
p = (coupled<double>)t;  // explicit cast: renormalize twofold to coupled
p = (coupled<double>)tf; // explicit cast: reshape and renormalize
```

Special functions for renormalizing with same basic type:

```
twofold<double> t;
coupled<double> p;
p = renormalize(t); // general case
p = fast_renorm(t); // special, if t.error does not exceed t.value
```

Auxiliary functions for taking value and error parts:

```
twofold<double> t;
coupled<double> p;
double x0, y1;
x0 = value_of(t);
y1 = error_of(p);
```

If argument x is dotted, then `value_of(x)` returns x and `error_of(x)` returns 0. This intends to support shape-generic coding, make same-looking code work for shaped and dotted numbers.

## Functions

C++ interface overloads functions like tadd/padd in two ways: removes suffix 'f' for C-style functions if `float` arguments, and defines functions of `twofold<T>` and `coupled<T>` arguments. For example:

```
// plain C style
float x0,x1, y0,y1, z0,z1;
z0 = tadd(x0,x1,y0,y1,&z1); // no 'f' suffix in function name
z0 = padd0(x0,y0,&z1);

// C++ style
twofold<float> x, y, z;
coupled<float> u, v, w;
float a, b;
z = tadd(x,y);
z = tadd(x,b); // twofold + dotted
z = tadd(a,y); // dotted + twofold
z = tadd(x,b); // twofold sum of dotted arguments
w = padd(u,v);
```

Overloading C-style functions does not support AVX packed types like __m256 et al. C++ style functions do not support AVX types either. My reasoning is that AVX extension is processor-specific, so functions for it should follow the AVX intrinsic style.

C++ style functions overloading allows using same name for all cases if some or all parameters dotted. Note however that prefix 't' or 'p' would explicitly distinguish twofold from functions coupled-precision. Following is full list of arithmetic twofold/coupled functions. Selection {t|p} is mandatory prefix:

| Function | Description |
|---|---|
| {t\|p}add | Add |
| {t\|p}sub | Subtract |

| Function | Description |
|---|---|
| `{t|p}mul` | Multiply |
| `{t|p}div` | Divide |
| `{t|p}sqrt` | Square root |

Special twofold functions compute multiply, divide, and square root operations faster as arguments are coupled-precision, with error part much smaller than value. Unlike plain C, types control automatically ensures arguments meet this pre-condition:

| Function | Description |
|---|---|
| `z=tmulp(x,y)` | Multiply |
| `z=tdivp(x,y)` | Divide |
| `z=tsqrtp(x)` | Square root |

Auxiliary functions renormalize twofold into coupled, and implement fast add/subtract operations:

| Function | Description |
|---|---|
| `p=renormalize(t)` | Renormalize twofold into coupled |
| `p=fast_renorm(t)` | Renormalize fast if argument's error is small |
| `p=fast_add0(x,y)` | Fast summation if dotted |x|≥|y| |
| `p=fast_sub0(x,y)` | Fast subtract x-y, if dotted |x|≥|y| |

C++ style interface defines functions for comparing twofold and coupled numbers. Comparing twofolds works like comparing their value parts, so if you replace all or some of numbers with twofolds in your code its behavior would not change. Comparing coupled numbers is more fine-grained: if value parts equal, we should compare error parts. Comparing with NaN results in `false` as usual for C/C++.

Full list of comparing functions. They accept twofold/coupled/dotted arguments and return `bool`:

| Function | Description |
|---|---|
| `{t|p}lt(x,y)` | Check if x < y |
| `{t|p}le(x,y)` | Check if x ≤ y |
| `{t|p}gt(x,y)` | Check if x > y |
| `{t|p}ge(x,y)` | Check if x ≥ y |
| `{t|p}eq(x,y)` | Check if x = y |
| `{t|p}ne(x,y)` | Check if x ≠ y |

Additionally, a few functions would serve for simple operations, specifically:

| Function | Description |
|---|---|
| `{p|t}neg` | Negate x.value and x.error |
| `{p|t}abs` | Absolute value: neg(x) if x.value<0 |
| `{p|t}isinf` | Check if x.value or x.error is infinite |
| `{p|t}isnan` | Check if x.value or x.error is NaN |

Functions looking like standard:

| Function | Description |
|---|---|
| `z=fabs(x)` | tabs(x) or pabs(x) depending on shape of x |
| `z=sqrt(x)` | tsqrt(x) or psqrt(x) depending on shape of x |
| `isinf(x)` | tisinf(x) or pisinf(x) depending on shape of x |
| `isnan(x)` | tisnan(x) or pisnan(x) depending on x shape |

Functions for shape-generic programming, looking same whether parameter x is dotted or shaped:

| Function | Description |
|---|---|
| `value_of(x)` | Returns x.value, or just x if x is dotted |
| `error_of(x)` | Returns x.error, or just x if x is dotted |

To conclude, full list of C-style functions, same as for plain C except no suffix 'f' in function names:

- Main list: {t|p}{add|sub|mul|div}[0|1|2] and {t|p}sqrt[0]
- Special list: tmulp, tdivp, tsqrtp
- Auxiliary: renormalize, fast_renorm, fast_add0, fast_sub0

Full list of C++ style functions, same except no '0', '1', '2' if dotted arguments, plus comparison:

- Main list: {t|p}{add|sub|mul|div} and {t|p}sqrt
- Special list: tmulp, tdivp, tsqrtp
- Auxiliary: renormalize, fast_renorm, fast_add0, fast_sub0
- Comparison: {t|p}lt, {t|p}le, {t|p}gt, {t|p}ge, {t|p}eq, {t|p}ne
- Service: {t|p}neg, {t|p}abs, {t|p}isinf, {t|p}isnan
- Looking standard: fabs, sqrt, isinf, isnan
- Shape-generic: value_of, error_of

## Operators

C++ interface defines arithmetic operators +x, -x, x+y, x-y, x*y, x/y accepting twofold arguments and calling appropriate twofold function. If one of arguments is "dotted" (not twofold or coupled-precision), operator calls appropriate twofold function. If basic types differ operator converts to common type, that is converts `float` to `double`, or `int` to `float` or `double`, or twofold of `float` to twofold of `double`.

If arguments are coupled, operator calls twofold operation anyway so result is twofold. If you still need coupled-precision arithmetic, use functions like `padd()` explicitly. Though I think in most cases, you may use twofolds and renormalize. In particular, following demonstrates twofolds for improving accuracy in dot product, see [Use examples](#)/[Dot-product](#), and XBLAS [7]:

```
twofold<T> s=0;
for (int i=0; i<m; i++)
    s += x[i]*y[i];
T result = s.value + s.error; // compensate accumulated rounding errors
```

C++ interface defines logical operators x<y, x<=y, x>y, x>=y, x==y, x!=y accepting twofold arguments. If one of arguments is dotted, operator calls twofold comparison. If basic types differ, operator converts to common. If arguments are coupled-precision, operator compares as twofolds anyway. This way, if you replace all or some numbers with twofolds in your program, program behavior must not change.

## Comparing

Comparing twofolds by value parts, ignoring error parts, is non-trivial design decision. Let me explain it:

Twofolds main purpose is debugging accuracy problems by detecting if calculation appears too sensitive to precision. Comparing twofolds, say $x = x_0 + x_1$ versus $y = y_0 + y_1$, is good chance for detecting, as $x_0$ and $y_0$ may compare differently than more accurate $x_0 + x_1$ and $y_0 + y_1$. For example, $x_0$ might be less than $y_0$, but $x_0 + x_1$ be greater than $y_0 + y_1$, or conversely. If such discrepancy, we might logically conclude that twofold comparing $x$ versus $y$ results in "indefinite".

Now consider an if-then-else operator with comparing twofolds as the branching condition, for example:

```
if (x < y)
    ...;
else
```

```
       ...;
```

If results of comparing twofolds "x < y" is indefinite, say value parts compare like `true` but comparing value + error combinations results in `false`, what branch is right for executing?

If comparing is indefinite, this probably means the standard 1x precision appears not enough for making correct branching decision. Such lack of accuracy is probably the bug in the program, exactly that sort of bugs which twofolds target. So probably right decision would be terminating the program, maybe throw an exception to bring programmer's attention to this situation.

However, I decided to process such situations silently, as aligns with processing NaN quietly in C/C++.

### Namespace

C++ interface defines almost all of its stuff inside the "tfcp" namespace. Exception is overloading plain C functions for accepting `float` type. That is C-like variant of `tadd()` et al belongs to default namespace:

```
float x0,x1, y0,y1, z0,z1;
z0 = tadd(x0,x1,y0,y1,&z1); // no suffix 'f' in function name
```

While variant of `tadd()` et al over C++ types belongs to "tfcp" namespace:

```
twofold<float> x,y,z;
z = tfcp::tadd(x,y); // using namespace "tfcp"
```

### Printing

For printing twofold and coupled numbers, I propose the following format, with error part in square brackets:

- Unity exactly: 1[0]
- π as twofold<float>: 3.14159[-8.74228e-08]

My experimental implementation defines operator "out << x" for printing twofolds:

```
twofold<T> x;
cout << "x: " << x << endl;
```

You may modify format as usual, print hexadecimal, change amount of significant digits, etc.

### To do list
- Functions
- Complex
- Vectors

### Functions

This C and C++ interface obviously lacks of elementary functions, sine/cosine et al. Twofold and coupled-precision variant of elementary functions is on my "to do" list.

### Complex

I see two ways for combining twofold with complex both having reasons:

- complex<twofold<T>> must work for compatibility
- twofold<complex<T>> might compute more accurately

Supporting complex numbers is another "to do" on my list.

### Vectors

High performance of valarray<twofold<T>> and maybe twofold<valarray<T>> is on my "to do" list.

For plain C, I think a few functions like vtadd() for vector twofold summation might be reasonable.

# Implementation

- [Static vs inline](#)
- [Standard C/C++](#)
- [AVX extensions](#)
- [Compiler flags](#)

In this section, I explain specifics of my experimental implementation. This is all about performance.

## Static vs inline

Both C and C++ interfaces implemented with same "`twofold.h`" header file. It consists of universal part applicable for both C and C++, followed by part specific for C++.

Twofold add, subtract, et al are technically very simple functions each involving several basic operations sequentially without branching. For example:

```c
static float pmul0f(float x, float y, float *r1) {
    float r0 = dmulf(x,y);   /* r0 = x*y           */
    float t  = dnegf(r0);
        *r1 = dfmaf(x,y,t);  /* r1 = fma(x,y,-r0) */
    return r0;
}
```

There are two tricks to explain here. First, I use `static` keyword instead of `inline` to make this code compatible with both C and C++. This works fine as modern compilers perfectly inline static functions. My testing with GNU and Microsoft compilers shows performance up to 100% (like 99% and more) of processor's capacity.

Second, I do not define linking, like `extern "C"`, for this and other plain C functions. So these function names mangling differ if in plain C or in C++ context. This allows overloading if in C++, so same function name would work for `float` and `double` parameter types, without the 'f' suffix in function name.

## Standard C/C++

For implementing pmul0f() we use functions like dmulf() which is nothing more than basic multiplying. The reason for defining functions like dmulf(x,y) instead of just writing x*y is that we provide two ways for implementing such basic operations: via standard C/C++ coding, and via AVX intrinsic functions:

```c
// C/C++ standard
static float dmulf(float x, float y) {
    return x*y;
}

// Using AVX intrinsic
static float dmulf(float x, float y) {
    return _mm_cvtss_f32(_mm_mul_ss(_mm_set_ss(x),_mm_set_ss(y)));
}
```

Implementation via AVX is default; to activate coding with standard C/C++ compile with –DNOAVX flag. Following are two reasons for such design:

First is about Microsoft compiler. The standard C/C++ library implements the FMA function very slowly. This causes twofold multiply, divide, and square root perform hundreds times slower, below 1 megaflop on my test computer, which is unacceptable. Implementing via AVX intrinsic fixes this problem. Though add/subtract may work faster if implemented via standard C/C++, I decided making AVX the default.

Second is about compiler optimizations. For high performance you probably compile with fast-math, GNU -ffast-math of Microsoft /fp:fast options. But optimizing for fast-math, compiler would replace tricky expressions like (x+y)-y with just x, so damage the twofold arithmetic which needs such tricky expressions for assessing rounding errors. Implementing via AVX intrinsic would prevent twofold functions from such damaging optimization, so you can compile with fast-math options.

## AVX extensions

By default, "`twofold.h`" does not expands twofold functions for AVX intrinsic. To define functions like `_mm256_tadd_pd()` et al, compile with –DAVX option.

## Compiler flags

The above Standard C/C++ and AVX extensions subsections explain the –DAVX and –DNOAVX options and compiling for strict or either fast math. Another important option is –mfma for GNU and /arch:AVX (or /arch:AVX2) for Microsoft compiler.

If compiling with Microsoft, you may omit /arch:AVX if compiling with /DNOAVX. In this case, twofolds would implement via C/C++ standard operations and library fma() and sqrt() functions. Twofold add and subtract operations would be fast, but multiply, divide, and square root would be desperately slow due to very slow fma()/fmaf() functions from Microsoft standard library.

Compiling with Microsoft, I recommend default or /DAVX options, with /arch:AVX or /arch:AVX2 which enable the Intel AVX intrinsic for the compiler.

If compiling with GNU, option –mfma is mandatory even if compiling with –DNOAVX. GNU standard library FMA functions may replace fma(x,y,z) with xy+z if compiler "thinks" fused-multiply-add is not supported by hardware. This makes twofold operations depending on FMA fast but useless, because twofold multiply, divide, and square root would not be able to estimate rounding errors.

The following table summarizes this data:

| GNU | Microsoft | Description |
|---|---|---|
| <default> | <default> | Implement dotted functions via AVX intrinsic |
| -DNOAVX | /DNOAVX | Implement dotted functions via standard C/C++ |
| -DAVX | -DAVX | Define AVX intrinsic-like functions for twofolds |
| -ffast-math | /fp:fast | May use if twofolds are default or with -DAVX |
| <default> | /fp:strict | Mandatory if compiling twofolds with -DNOAVX |
| -mfma | /arch:AVX /arch:AVX2 | Mandatory for GNU, optional for Microsoft (required if compiling with default or /DAVX) |

For best performance of scalar code, compile with NOAVX for GNU either with default for Microsoft:

```
g++ -mfma -DNOAVX …
gcc -mfma -DNOAVX …
cl /arch:AVX /fp:fast …
```

If manually vectoring your code for AVX, compile with –DAVX option:

```
gcc -mfma -ffast-math -DAVX …
cl /arch:AVX /fp:fast /DAVX …
```

You may observe "right" use of compiler options if you look at the make files I provide with twofolds code and tests/examples. Available at my web site [2], download code.zip archive.

# Software

- [Download](#)
- [License](#)

## Download

This article and C/C++ code available at my web site: <https://sites.google.com/site/yevgenylatkin/>

At this web site, see the page dedicated to this article: Twofolds in C and C++. Download text as PDF, experimental implementation source codes and testing logs within code.zip archive.

If any question or idea, like a change to interface, please contact me. Email: yevgeny.latkin@gmail.com

## License

The license for this article and its accompanying software, twofolds implementation and tests, is "free for any sort of academic and non-commercial use".

This includes learning, teaching, testing, using for testing your math software, and creating your own implementation. But please do NOT change interface! In you create your own implementation, please make interface compatible. If you have any proposal on changing interface, please contact me. Let us coordinate interfaces to prevent incompatible versions.

About possible commercial use of my implementation, I am afraid current version is not good enough.

# Conclusion

Strange fundamental fact about floating-point is that it works usually better than it theoretically should.

Rounding errors tend to compensate each other in typical computations so accumulated inaccuracy uses to grow moderately. Mathematic software usually counts on such typical computations; and works very well, usually, until non-typical case happens with unpleasant or even catastrophic consequences. In such a situation, we normally analyze how we should have predicted it.

I am not sure we should predict. I would not expect an "ordinary" programmer be so smart to predict everything. I think future computers should automate predicting, become smarter, automate generating software and proving its correctness, maybe use interval methods for guaranteeing boundaries. Smarter computers is long and difficult way; twofolds is lower-hanging fruit that we can leverage right now.

Strange fundamental fact about twofolds is that they also work much better than theoretically should.

Identifying problems is fundamentally easier task than fixing. Even if twofold precision is not enough for solving accurately, twofolds often can detect solution sensitivity to precision. So twofolds can increase confidence in math software. Suppose 90% of computations be "typical", and twofolds detect problems in 90% of non-typical cases. So result becomes 99% confident: either is accurate, or twofold catches its inaccuracy. Unfortunately, 1% remains uncaught, and we still may meet unpleasant consequences.

Twofolds cannot guarantee 100% confidence, but I think this is step in the right direction. I am trying to push this step with this series of articles about twofolds.

With this article, I propose C/C++ interface for twofold arithmetic, and experimental implementation for GNU and Microsoft compilers. I am trying to prove the concept of assuring quality of math software by testing its accuracy with twofolds. Twofolds high performance must allow testing permanently, on-fly, tracking inaccuracy accumulation in parallel with main computations.

Performance testing shows this may be technically possible even with my experimental implementation, which is less than 2000 lines of code in standard C and C++ with processor-specific intrinsic for Intel AVX. Defining twofold arithmetic as set of `static` functions in "`twofold.h`" allows leveraging up to 100% of processor potential with modern C/C++ compilers, though if vectoring for AVX manually.

I propose standardizing C and C++ interfaces for twofold arithmetic, and improve implementation in future compilers, and maybe processors. Compilers might support automatic vectoring algorithms for SIMD (like AVX), which would be easy once compiler gets educated about twofolds. Hardware might support faster operations for estimating rounding errors, details in [1].

Meanwhile, even current version of twofolds for C++ must be enough for investigating math algorithms. In many cases, I think twofolds would be the only practically applicable instrument for controlling errors. Are such cases widespread? Do twofolds really help assuring/debugging? Is instrumenting with twofolds worth investments? Answering honestly requires further investigation.

With this article, I intend to convince you trying. My experimental implementation is free for any sort of academic and non-commercial use; but please agree with me if you propose changing interfaces. Please do not hesitate to propose and/or ask questions. My email: yevgeny.latkin@gmail.com

Obvious gaps in current version of twofolds are lack of elementary functions (sine/cosine, et al), support for complex data, and fast array operations. This is on my "to do" list.